\documentclass{article}
\usepackage{arxiv}

\usepackage{hyperref}       
\usepackage{url}            
\usepackage{booktabs}       
\usepackage{amsfonts}       
\usepackage{nicefrac}       
\usepackage{hyperref}
\usepackage[T1]{fontenc}
\usepackage{microtype}      
\usepackage{lipsum}
\usepackage{graphicx}
\graphicspath{ {./images/} }

\usepackage{slashbox}
\usepackage[backend=biber,style=authoryear]{biblatex}

\usepackage{booktabs} 
\usepackage{color,soul}
\usepackage{fancyvrb}
\usepackage{amsmath}
\usepackage{amsthm}
\usepackage{float}
\usepackage{bbm}

\usepackage{graphicx}
\usepackage{lipsum}
\usepackage{xfrac}
\usepackage{colortbl}
\usepackage{multirow}
\usepackage{balance}
\usepackage{paralist}
\usepackage{tabularx}
\usepackage{etoolbox}
\usepackage{caption}
\usepackage{subcaption}
\usepackage{algorithm}
\usepackage{algpseudocode}
\usepackage{grffile}
\usepackage{enumitem}

\let\chapter\undefined 

\fvset{frame=single,framesep=1mm,fontfamily=tt,numbers=left,framerule=.3mm,numbersep=1mm,commandchars=\\\{\}}
\definecolor{bgred}{RGB}{255,210,205}
\definecolor{bgblue}{RGB}{210,220,255}
\definecolor{bgyellow}{RGB}{255,255,109}
\definecolor{bggrey}{RGB}{223,223,225}
\definecolor{bgpink}{RGB}{255,202,239}
\definecolor{bgsky}{RGB}{182,219, 255}
\definecolor{purple}{RGB}{180,0,180}

\definecolor{hvygreen}{RGB}{0,128,0}
\definecolor{hvypink}{RGB}{102,0,51}

\newcommand{\var}[1]{\operatorname{\mathit{#1}}}

\renewcommand{\epsilon}{\varepsilon}
\renewcommand{\phi}{\varphi}

\newcommand{\textcd}[1]{{\footnotesize \textup{\textsf{#1}}}}

\newcommand{\algoname}[1]{\textcd{#1}}
\newcommand{\dataset}[1]{\var{#1}}

\newcommand{\eat}[1]{}

\title{Learning Approximation Sets for Exploratory Queries}

\author{
 Susan B.Davidson \\
  University of Pennsylvania \\
\texttt{susan@seas.upenn.edu} \\
   \And
 Tova Milo \\
  Tel Aviv University \\
  Tel Aviv, Israel \\
\texttt{milo@cs.tau.ac.il} \\
  \And
 Kathy Razmadze \\
  Tel Aviv University \\
  Tel Aviv, Israel \\
\texttt{kathyr@mail.tau.ac.il} 
    \And
 Gal Zeevi \\
  Tel Aviv University \\
  Tel Aviv, Israel \\
\texttt{galzeevi@mail.tau.ac.il} 
}
\begin{document}
\maketitle
\begin{abstract}

In data exploration, executing complex non-aggregate queries over large databases can be time-consuming. Our paper introduces a novel approach to address this challenge, focusing on finding an optimized subset of data, referred to as the \emph{approximation set}, for query execution. The goal is to maximize query result quality while minimizing execution time. We formalize this problem as Approximate Non-Aggregates Query Processing (ANAQP) and establish its NP-completeness. To tackle this, we propose an approximate solution using advanced \emph{Reinforcement Learning} architecture, termed \algoname{ASQP-RL}. This approach overcomes challenges related to the large action space and the need for generalization beyond a known query workload. Experimental results on two benchmarks demonstrate the superior performance of \algoname{ASQP-RL}, outperforming baselines by $30\%$ in accuracy and achieving efficiency gains of $10-35X$. Our research sheds light on the potential of reinforcement learning techniques for advancing data management tasks. Experimental results on two benchmarks show that
\algoname{ASQP-RL} significantly outperforms the baselines both in terms of accuracy (30\% better) and efficiency (10-35X). This research provides valuable insights into the potential of RL techniques for future advancements in data management tasks. 
\end{abstract}


\section{Introduction}
\label{sec: intro}


In the data exploration phase of data science, users execute a variety of queries, including both aggregate and complex non-aggregate queries, often in an iterative manner~\cite{ATENAUserStudy}. However, when the database is large and the queries are complex, calculating exact answers for exploratory queries can be time-consuming, prompting users to prefer faster but approximate query results. Approximate Query Processing (AQP) \cite{AQPSurvey} has been proposed to accelerate the processing of aggregate queries, but does not address the complex, non-aggregate, SQL queries that are common in data exploration. 

For example, a recent study of the IMDB Database~\cite{JOB} showed that in the query workload of data exploration sessions, only about $50\%$ of the queries were aggregate queries.  The remaining half were complex select-project-join (SPJ) queries. These complex queries had an average processing time of 25 minutes, and $30\%$ of them exceeded 1 hour, even with result limits being imposed, making the wait time impractical. 

This paper extends the scope of approximate query processing to non-aggregate queries, presenting a solution that accelerates the execution of such queries while providing high accuracy.

\paragraph*{\textbf{Approximating complex non-aggregates queries}} Our work focuses on (1) defining what a reasonable approximation is for complex non-aggregate queries, and (2) providing an efficiently computed subset of the data, called an \textit{approximation set}, for evaluating complex non-aggregate queries. While approximate answers are well understood for aggregate queries, extending this idea to non-aggregate queries requires careful consideration. 
Our premise is that an approximation for a non-aggregate query should encompass some, but not necessarily all, tuples in the query result. This stance is grounded in practicality: when the query output becomes excessively large for human comprehension, users may not discern that the result is approximated. 

Unlike in AQP, where the quality metric for approximation remains consistent across queries, the critical factor in non-aggregate queries is the portion of the result that a user can thoroughly analyze. In scenarios where the user can analyze the entire result, the approximation must be highly accurate, covering the entire result set. However, in queries with exceptionally large results, the approximation should provide the user with an understanding of the query answer without necessarily containing more than the user can observe. As with aggregate queries, complex non-aggregate queries can take a long time to execute. Drawing parallels to users' willingness to accept extended, offline pre-processing times for fast, accurate approximate query processing in AQP, we posit a similar situation holds for our scenario: Users are willing to accept an extending pre-processing time that calculates a good approximation set which can then be used to produce fast, high quality approximate answers for queries. However, there are a number of challenges that must be addressed in order to find a good approximation set.

\paragraph{\textbf{Challenges}} Selecting an approximation set for subsequent querying presents various challenges: (C1) The vast combinatorial space, created by the number of combinations of tuples from different tables. (C2) Understanding the search space, due to the cost of executing the complete query workload.
(C3) The varying size of query results, which introduces different importance levels for tuples, e.g. tuples from queries with smaller result sizes are more significant than ones from larger result sizes. (C4) Generalizing for future, potentially dissimilar queries, since the query workload only captures what is already known.  (C5) Detecting when there is a drift in user interest. 

\paragraph*{\textbf{Inadequacy of Existing Approaches}} Despite the perception that existing solutions may address these challenges, there are several shortcomings. Online Analytical Processing (OLAP), for example, demands significant storage resources for data cubes, which can be up to ten times the size of the original dataset. OLAP also relies on proficient Database Administrators (DBAs) who are familiar with its unique syntax, limiting its accessibility for general users (see Section \ref{sec:related}). While AQP solutions seem promising, recent approaches using generative models (e.g., \cite{VAETABULAR}) often produce false tuples when applied to non-aggregate queries. This deficiency is critical when presenting non-aggregate query results, since false tuples may mislead users and lead to attempts to query irrelevant data. Other AQP works which create data subsets for subsequent model training  (e.g., \cite{verdictdb, Dbest}) also prove inadequate for non-aggregate queries (comparisons with AQP baselines are given in Section \ref{sec:exp}). Classical methods, including diverse sampling techniques and caching in managed database systems, are also considered in Section \ref{sec:exp}, but do not produce satisfactory results for non-aggregate queries.

\paragraph{\textbf{Our solution}}
We offer an innovative approach for efficiently approximating non-aggregate query results. Given an initial query workload and database (which could be multiple tables), our method involves creating a subset of the data, called an approximation set, to enable faster queries when the runtime for executing queries on the full database is excessive. To gauge an approximation set's quality with respect to the query workload, we define a metric. Unfortunately,  optimizing with respect to this metric is NP-hard, so we turn to Reinforcement Learning (RL) for an approximate solution. 

Our system, ASQP-RL, initiates a comprehensive data and query workload pre-processing phase, resulting in an initial reduced subset of the data (C2). This subset is translated into the RL action space, where, at each training step, the model predicts the next tuples to include in the approximation set based on the current state. This predictive action aims to maximize the reward, representing the quality of results for queries encountered so far. ASQP-RL employs a reinforcement learning model utilizing critic-actor networks with Proximal Policy Optimization (PPO), specifically tailored for tabular data. The model encompasses a well-crafted environment and transformed action definitions (C1, C4). A specialized reward function that aligns with the goal of selecting rows for accurate approximations (C3). This combination yields a robust, adaptive solution that significantly improves the efficiency and accuracy of non-aggregate query approximations. Additionally, our system is equipped to detect drift in user interests, enabling model fine-tuning for better approximate results (C5).

While not explicitly addressed, our approach can also handle aggregate queries in the query workload by rewriting them to SPJ queries. Users can subsequently include aggregate queries in their data exploration sessions and obtain surprisingly good results, as shown in Section \ref{sec:exp}.

\paragraph*{\textbf{Benefits}} Our solution offers several benefits: 
\begin{enumerate}
    \item  A versatile mediator adaptable to any database, enhancing flexibility across diverse environments.
    \item Flexible training within time constraints, providing estimated guarantees and managing the tradeoff between running time and result quality.
    \item Efficient query interpretation through vector-based techniques, enabling a rapid reduction of the initial space, facilitating swift training processes and contributing to overall system efficiency.
    \item Incorporating a predictive model and query interpretation assesses the likelihood of missing answers, enhancing result accuracy.
    \item The ability to detect interest drift, enabling fine-tuning to evolving exploration needs. This is achieved through the generation of initial query workloads and dynamic fine-tuning of the model with adapted queries. 
\end{enumerate}
These benefits position ASQP-RL as a comprehensive tool for efficient and effective data exploration.

\paragraph*{\textbf{Contributions and Outline}}
In addition to providing a comprehensive tool for data exploration, this work is among the first applications of advanced RL concepts.  Specific contributions include:
\begin{enumerate}
\item \textbf{Problem Definition and Metric Introduction} (Section \ref{sec:problem_def}): We define the problem of approximating non-aggregate queries and introduce a novel metric for evaluating the quality of approximations.
\item \textbf{Comprehensive RL-Based Solution} (Sections \ref{sec:system_framework} and \ref{sec: rl_in_asqp_rl}): We present a holistic solution based on RL, specifically ASPQ-RL. This includes the training of a model to create an approximation set. Additionally, we address challenges such as determining when to use the approximation set for answering queries, when to fine-tune the model in response to a drift in user interests, and how to handle an unknown query workload.

\item \textbf{Experiments Demonstrating Efficiency and Quality} (Section \ref{sec:exp}): We conduct comprehensive experiments showcasing the efficiency and quality of ASQP-RL for non-aggregate queries. The results highlight its adaptability to diverse time constraints and superior performance in managing the trade-off between running time and result quality.
\end{enumerate}
Section~\ref{sec:related} presents related work, and Section~\ref{sec:conc} concludes the paper.

\section{Related work}\label{sec:related}
We provide an overview of key areas closely related to our study, categorizing them into five groups. 
These fields collectively provide a comprehensive understanding of existing literature and highlight the unique contributions of our work.

\paragraph{OLAP} In Online Analytical Processing (OLAP), especially within Multidimensional OLAP (MOLAP), the focus is on aggregate operators managed through data cubes relying on past query workloads (\cite{olapSurvey, olapSurvey2}).However, MOLAP poses several challenges: Spatial requirements for these cubes, highlighted in \cite{covidOlap}, are substantial, about ten times the original dataset size, imposing significant storage overhead; Implementing MOLAP systems demands DBAs with specific OLAP syntax expertise, limiting accessibility to average data scientists\cite{IBMOlapSyntax}; The setup process, per \cite{IBMOlapSetup}, is intricate, potentially taking several days. These limitations reduce OLAP tools' accessibility in the data science community, making them less suitable for our specific problem.

\paragraph{Approximate Query Processing (AQP) and Generative Models} In AQP, two main approaches aim to provide approximate answers to aggregation queries(\cite{AQPSurvey}): \textbf{(1)} \emph{Retaining a small number of tuples with metadata}: such as recent sampling methods for this purpose \cite{apqKraska2017, cormode2017data, agarwal2013blinkdb}, aim to expedite query processing by reducing tuples at the cost of exactness. While those methods are less effective for non-aggregate queries in our specific use case (Section \ref{sec:exp}), we draw inspiration from \cite{verdictdb} to generate an initial subset of joined tuples from relevant tables in the query workload. This subset forms the basis for training a model that produces a high-performing subset for non-aggregate query approximation tasks. \textbf{(2)} \emph{Generating new tuples using generative models}: Generative models, such as generative adversarial networks (GANs) and Variational Autoencoders (VAEs), have been extensively explored for generating synthetic or representative samples in the context of tabular data \cite{GANTABULARSURVEY, VAETABULAR}. These models offer advantages like explicit probabilistic modeling and capturing complex data distributions. However, as demonstrated in Section \ref{sec:exp}, generated tuples may significantly deviate from real tuples, leading to issues where user queries produce minimal or no results, return nonsensical outcomes, or even mislead the user. This makes them unsuitable for our particular case.

\paragraph{Data Summarization and Sampling}
These techniques provide essential tools for various data processing tasks. Cluster-based representative selection aims to maximize data diversity by selecting instances from clustered data \cite{usingTreeForest}. In visualization, row sampling reduces data points while minimizing error \cite{VisualizationAwareSampling}. Query result diversification selects rows with both relevance and diversity, often utilizing greedy algorithms \cite{queryResultDivers}. However, as demonstrated in Section \ref{sec:exp}, these approaches, optimized for different objectives, are unsuitable for our specific use case.

\paragraph{Data Reduction, Sketches and others}
Data sketches and reduction techniques, including HyperLogLog and Count-Min Sketch \cite{cormode2017data, ur2016big}, are impactful for data processing tasks. However, they may lose critical data details needed for accurate selection query results, making them unsuitable for our purposes. Approaches like the skyline operator~\cite{skyline}, caching operations~\cite{caching}, and view selection materialization with space constraints (\cite{viewSelectionMat}) are relevant but prove inferior for our use case, as demonstrated in our experiments.

\newcommand{\probP}{\text{I\kern-0.15em P}}
\newcommand{\probE}{\mathop{\mathbb{E}}}
\newcommand{\indicator}{\mathbbm{1}_{\{t\in q_i(T)\}}}

\section{Problem Definition and Complexity}\label{sec:problem_def}
We now introduce the problem of Approximate Non-Aggregate Query Processing (ANAQP). 
We begin by defining the problem in a simplified scenario with a known query workload, and give a metric to assess the quality of a set of subsets of tuples with respect to the given workload. 
In the next section we extend these concepts to scenarios where the query workload is unknown. 

\paragraph{ANAQP Problem Definition} 
Consider a set of relational instances $\mathcal{T}= \{T_1, ..., T_n\}$
and query workload $Q$
consisting of select-project-join (SPJ, non-aggregate) queries over $\mathcal{T}$.
We denote $w$ as a function $w:Q \rightarrow [0,1]$ that assigns a weight to each query, satisfying the constraint $\sum_{q\in Q}{w(q)} = 1$.

We assume limits on the size of the available memory ($k$) as well as on the maximum result size that users can cognitively process (frame size $F$).
In practice, $F$ may vary from 10 rows (the default data view size in pandas) to 500 rows (the default result size in PostgreSQL), and is configurable.
\eat{$F$ is different from the parameter $k$, which is the maximum sample size for the input dataset, since it relates to the maximum size of the output for human cognition.}

\paragraph*{Metric}
Our metric function, $score(S)$,  measures the quality of a set of subsets of the relational instances in $\mathcal{T}$, $\mathcal{S}= \{S_1, ..., S_n\}$ where $S_i\subseteq T_i$, with respect to $(\mathcal{T},Q,w,k,F)$:
\begin{equation}\label{eq:metric}
score(\mathcal{S}) = \frac{1}{|Q|}\sum_{q\in Q}{w(q)\cdot min(1, \frac{|q(\mathcal{S})|}{min(F,|q(\mathcal{T})|)})}
\end{equation}
Intuitively, if a query returns more tuples than the frame size $F$, there is no need to include in $S$ more than $F$ tuples from the query result since the user may not be able to cognitively process them. 
Conversely, if a query returns only a few tuples, then each tuple carries significant importance in the result. The metric takes into account the weight assigned to each query and calculates the average score over all queries in $Q$.

\paragraph*{ANAQP Goal}
Given an instance $\mathcal{T}$, a set of queries $Q$, weight function $w$, and positive integers $k$ and $F$,  the goal of ANAQP is to find a set $\mathcal{S}$ of subsets of tables in $\mathcal{T}$ where $\sum_{S_i \in \mathcal{S}} |S_i| < k$
such that $score(\mathcal{S})$ is maximized.

\paragraph{Problem Hardness}\label{subsec:hardness}
The ANAQP problem is already hard when queries are over a single table, $n=1$.
In this case, a naive approach to is to consider all subsets $S \subseteq T$ of size $k$, and find the one that maximizes  $score(S)$.
This is obviously computationally intractable.  Worse yet, this naive solution cannot be significantly improved since the ANAQP problem is NP-hard. This can be shown by reduction to the max-$k$-vertex cover problem \cite{maxkcoverHardness}. We create an ANAQP instance from the input to the max-$k$-vertex cover problem as follows:
For each vertex $v\in V$, we create a tuple  $t_v \in T$, and for each edge  $e=\{u,v\}\in E$, we create a query $q_e(T) =\{t_u, t_v\}$ that includes the tuples corresponding to the edge's endpoints $q_e(T) =\{t_u, t_v\}$. The query weights are set to the weights of the corresponding edges in the graph. The limit on the number of informative tuples, $F$ is set to $1$, and the memory size is set to $k$ (the input to the max-$k$-vertex problem). 
Observe that a solution to the ANAQP formulation of the problem is a solution to the max-$k$-vertex problem, as the score function $score(S)$ defined in Equation \ref{eq:metric} is equivalent to the objective function of the max-$k$-vertex cover problem.

In the case where multiple tables are involved ($n>1$) the problem is combinatorially even worse, since all possible combinations of subsets of each relation must be considered.

\paragraph*{Parameter Settings}
As can be seen from the problem formulation, the choice of parameter settings significantly impacts the overall performance of a solution: 
The frame size ($F$) directly affects the score function. The available memory ($k$) determines how large the subsets can be and therefore affects the accuracy and coverage of query results. 
Increasing $k$ enables a larger number of tuples to be kept, leading to a higher likelihood of obtaining exact query outputs and improved coverage for a wide range of queries. We show the effectiveness of our system under different parameter settings in Section \ref{sec:exp}.

\paragraph*{Aggregate Queries}

Aggregate queries are important in data exploration but are not directly addressed in our problem formulation, which focuses on SPJ queries.
Our approach to handling aggregate queries in the initial query workload is to transform them into SPJ queries over $T$ by removing the aggregate and group operators. 
We show in Section~\ref{subsec: AQP} that this naive approach yields surprisingly good results when compared with Approximate Query Processing techniques.

\eat{
It is also important to consider changes to other parameters such as the given tables ($T$). These tables may undergo regular updates or stream new data, introducing potential changes to the overall system behavior. Similarly, the query workload ($Q$) may evolve, encompassing new queries or alterations in the existing query distribution. Although these parameter changes lie beyond the scope of this work, they provide avenues for future research and exploration.
}

\paragraph{Unknown Query Workloads}
We have introduced ANAQP using a simplified scenario with a known query workload. However, the query workload is typically a subset of all possible user queries -- users continually issue new queries which may differ from the original workload. Even worse, the query workload may be unknown.  We address this problem in Section \ref{subsec:preproccessing}.

\vspace{0.1in}
Due to the hardness of finding an exact solution to ANAQP, as well as the cost of executing queries over large relations to evaluate $score(\mathcal{S})$, we show in the next section how to use Reinforcement Learning (RL) to find a good approximate solution.

\section{Approximation using RL} \label{sec:system_framework}
In this section, we outline our approach to employing Reinforcement Learning (RL) for Approximate Non-Aggregates Query Processing (ANAQP). We introduce the ASQP-RL framework, which serves as our solution for approximate non-aggregate query processing using RL (Section \ref{subsec:challenges}). The framework is subsequently detailed, covering the preprocessing steps for input tables and query workloads, along with addressing scenarios without a known query workload (Section \ref{subsec:preproccessing}). High-level descriptions of the RL training and inference steps are provided in Sections~\ref{subsec: training} and \ref{subsec: inference}. Further details of the RL components are deferred to Section \ref{sec: rl_in_asqp_rl}. Lastly, we discuss additional enhancements, including a light version of the system for improved running times and the scenario where no queries are initially available (Section \ref{subsec:unkown_q}).

\begin{figure*}
    \includegraphics[width=\textwidth]{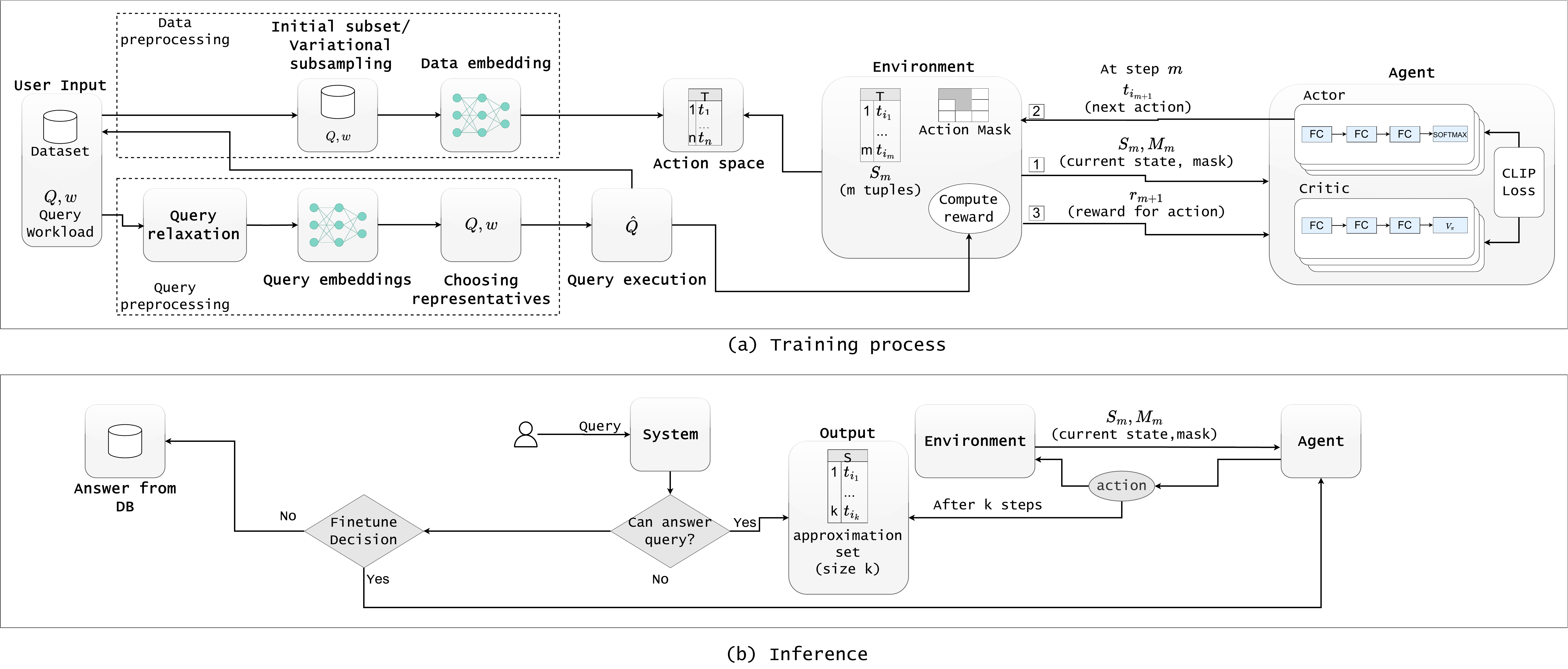}
\centering    
    \caption{ASQP-RL architecture.}
    \label{fig:rl_reduction_architecture} 
    \vspace{-4mm}
\end{figure*}
\subsection{Overview of \algoname{ASQP-RL}}\label{subsec:challenges}
In our RL approach, an approximation set consisting of tuples from the input tables is learned through trial-and-error interactions.
The RL agent starts with some initial state (approximation set), and at each step predicts the next action (tuples selection) based on the current state.
This predictive action aims to maximize the reward, representing the quality of results for queries encountered thus far (Equation~\ref{eq:metric}).
The use of RL ensures the fitness of selected tuples to the queries, while exploration introduces diversity in tuple selection.

The workflow of \algoname{ASQP-RL} is presented in Figure \ref{fig:rl_reduction_architecture}.  The training phase shown in (a) begins by pre-preprocessing the database $D$ and the query workload $Q$ to address some of the challenges mentioned earlier, and provide the input to the RL model. The RL model is then trained.
In the inference phase shown in (b), the framework generates an approximation set  using the trained model. 
When a query is issued by the user, the system decides whether to use the approximation set or query the full database, based on an estimation of how close the query is to those used to train the model.  


\subsection{Data and Query Pre-processing}\label{subsec:preproccessing}
The pre-processing phase depicted in Figure \ref{fig:rl_reduction_architecture}(a) prepares the data and queries for input to the RL phase and addresses several challenges: the combinatorial size of the solution space ($C1$),
and the high cost of executing the query workload ($C2$).  
We describe how queries and data are pre-processed and how these challenges are addressed in the context of the RL framework.  We then discuss how to generalize for future queries ($C4$).

\paragraph{Query Pre-processing}
Queries are first generalized before creating their vector representations. For this, we use existing \emph{query relaxation} methods \cite{query_relaxation}, which loosen or modify query conditions when strict adherence to the original conditions do not yield sufficient results. In our setting, these modifications enlarge the result set for each query, thereby making the query more general.  This also adds in tuples that are beyond what is returned by the current workload, helping to generalize for future, not yet known, queries. 
Vector representations of the generalized queries are then created using a modified version of $\algoname{sentence-BERT}$~\cite{sentence-BERT}. We adapt the transformer-based model to capture similar queries by embedding queries into a vector space. 
Finally, we use a clustering algorithm over the embedded queries to select what we call the \textit{query representatives} to be used in the score function. 

\paragraph{Data Pre-processing} 
Data pre-processing starts by removing data that is not used in any query representative. This is done by executing the query representatives over the full dataset and keeping only their output.
The output is further reduced through subsampling in order to create the action space to be used for training the model.
This reduced subset of tuples (the action space) is then transformed into a vector representation to be ingested by the RL model and to ease the learning. The subsampling technique used for reducing the results of the executed query representatives is \emph{variational subsampling} \cite{verdictdb}, a technique that uses probabilistic models, particularly in Bayesian inference, to approximate intractable likelihoods by introducing latent variables and optimizing their variational parameters. The transformation into a vector representation is achieved using another modified version of sentence-BERT. This version is designed to be more suitable for tabular rows, incorporating modifications such as including column names as tokens to capture both the meaning of the column as well as the value.

\paragraph{Addressing challenges $C1$, $C2$ and $C4$}
In the RL framework, challenge $C1$ becomes one of the size of the \emph{action space}. This space constrains the RL algorithm to valid tuple selections aligned with table structures, data distribution, and the provided query workload. Notably, selecting tuples separately from each table may lead to unjoinable tuples (\cite{verdictdb, quickr}). In our approach this is addressed in the data pre-processing phase, which creates a smaller set of tuples based on the generalized query workload, which may contain joins. 
The second challenge, $C2$, is the cost of executing queries in the query workload, in particular to evaluate the quality of an approximation set. This is addressed during the query pre-processing phase by choosing a smaller set of query representatives, and during the data pre-processing phase by selecting a smaller dataset over which to execute the queries.  To overcome the challenge of a future, unknown workload ($C4$), we use a fine-tuning strategy  (detailed next). Moreover, the use of query relaxation introduces more distinct tuples into the action space and avoids overfitting to the given query workload.  The RL builtin exploration phase also makes it return a more diverse solution,  further addressing this challenge.

\subsection{Training}\label{subsec: training}

The RL formulation of our problem treats it as a sequential decision-making process. The RL components during the training phase of \algoname{ASQP-RL}, depicted in Figure \ref{fig:rl_reduction_architecture}(a), are defined as follows.

\paragraph{Action Space} The database is transformed into an action space which serves as input for the RL model during training. Training involves formulating a policy for selecting actions from this space based on a given state. Since it is infeasible to consider all subsets of tuples across all tables, we create a significantly reduced action space as described in Section~\ref{subsec:preproccessing}.

\paragraph{State and Environment} The state, a crucial element capturing relevant information for the decision-making policy in Reinforcement Learning (RL) models, in our case is represented as the set of actions previously selected by the model. Recall that an action, in this context, encompass multiple tuples sourced from different tables. Thus, the state effectively encapsulates the approximation set chosen thus far. The environment, serving as the external system or context with which the RL agent interacts, defines how states are represented and how the action space is made available to the agent for policy definition. Specifically, we formalize the \algoname{ASQP-RL} environment in \textit{GSL} form, a description further elaborated upon in Section \ref{sec: rl_in_asqp_rl}. In that section, an alternative approach is also discussed, which, due to its limitations, was not incorporated into the algorithm. Nonetheless, the results of this alternative approach are presented in Section \ref{sec:exp}.

\eat{\paragraph{State and Environment} The state

is the set of actions previously selected by the model, and encapsulates the approximation set chosen thus far. The environment, serving as the external system or context with which the RL agent interacts, defines how states are represented and how the action space is made available to the agent for policy definition. In Section \ref{sec: rl_in_asqp_rl} we formalize the \algoname{ASQP-RL} environment.}

\paragraph{Reward} The reward, furnished by the environment, plays a pivotal role in updating the policy network. 
Our reward function is the score shown in Equation \ref{eq:metric}, which is defined by the current batch of queries.
To expedite the training process and foster stability in the learned policy, we implement a common technique of batch training and loss computation. Each epoch in the model training uses a distinct batch of queries, directly influencing the reward function. In essence, a high reward signifies that the chosen action (representative of several tuples) encompasses the majority or even all the results from the current batch of queries.

\paragraph{Agent} The agent considers the current state and then selects an action based on the learned policy, $\pi_\theta$. 
In \algoname{ASQP-RL}, our agent uses the actor-critic method \cite{A3C}. This method incorporates several actor networks and a critic network operating asynchronously. The actor networks are used to choose actions based on $\pi_\theta$, while the critic networks update the policy based on the received reward. In Section \ref{sec:exp} we conduct an ablation study, justifying each component of the agent in our system.

\paragraph{Overall training process}
The training process, shown in Figure \ref{fig:rl_reduction_architecture}(a) and detailed in Algorithm~\ref{alg:training}, takes as input the database $D$ and the query workload $Q$, which are first pre-processed. In this phase, queries from the training workload are generalized and prepared for evaluation (lines $1-2$). The results of these modified queries (line $3$) are then used to construct the action space $A$ (line $4$). The training step begins with the iterative selection of the approximation set using the RL model. Each epoch of the training process processes batches of queries retained from the pre-processing stage (line $6$), evaluating the chosen action and subsequently updating the model (lines $7-10$).

Specifically, each state denoted as $s_k$ represents the state after $k$ epochs of the model, indicating that $k$ actions were previously selected from the action space $A$. The agent, represented by the actor network, takes $s_k$ as input and computes the probability of each action, denoted as $\pi_\theta(a|s) = p(a| s,\theta)$, where $a \in A$. A higher probability indicates a larger expected long-term reward for the corresponding action. This implies that the final approximation set is more likely to satisfy the query workload the model was trained upon. Multiple actor-critic networks operate asynchronously during this process to expedite training by considering more policies.

Given the fact that actions represent tuples, we use action masking \cite{action_masking} to enforce the constraint that a tuple may only enter the approximation set once. In this technique, the environment provides the model along with the state a mask that shows the valid actions the model can choose from.

The critic network estimates the long-term reward of the current state, and together with the returned reward, computes the loss error (see Section \ref{sec: rl_in_asqp_rl} for more). The two networks are then updated for optimized action selection.

\paragraph*{Addressing Challenges $C3, C4$} In the RL context, challenges $C3$ (diversity and balance in results) and $C4$ (generalization for future queries) necessitate a well-crafted reward system and an effective exploration strategy. $(C3)$ is addressed by formulating a reward system that carefully balances diversity and relevance. The reward associated with selecting an action for a query with a modest result size holds significant value, emphasizing a balance between diversity and ensuring relevance. This reward mechanism is crucial during the exploitation phase, guiding the selection of actions for the creation of the approximation set. To tackle Challenge $(C4)$, ensuring selected tuples generalize effectively for future unseen queries, we employ a policy-based RL framework. This framework systematically chooses actions, incorporating an exploration strategy. The exploration strategy aims to explore diverse combinations of actions that are potentially relevant not only to the queries in the current workload but also to future queries. This approach enhances the model's ability to generalize effectively, accommodating a broader spectrum of queries beyond the ones encountered during training.

\eat{\paragraph*{Addressing Challenges $C3, C4$} In the context of RL, the $C3$ challenges (\emph{diversity and balance in results}) is reflected by the presence of varying query result sizes that requires a careful design of the \emph{reward} to guide the selection of tuples. In particular, it is important to strike a balance between diversity and ensuring relevance for queries with distinct result sizes. For instance, the reward associated with selecting an action pertinent to a query with a modest result size may hold greater value than an action relevant to a query with a very large result size. This reward mechanism is a pivotal component of the \emph{exploitation} phase, facilitating the identification of actions that are most relevant to the creation of the approximation set. Next the challenge $C4$ (\emph{generalization for future queries}) is reflected as ensuring the selected tuples generalize effectively for future unseen queries. To achieve this objective, we leverage a policy-based RL framework. This framework systematically chooses actions, employing an \textit{exploration} strategy that explores diverse combinations of actions that are potentially relevant both to the queries in the query workload as well as to future ones.} 

\begin{algorithm}
\caption{\algoname{ASQP-RL Training}}\label{alg:training}
\begin{algorithmic}[1]
\Require Database $D$, query workload $Q$
\Ensure A trained \algoname{RL} model
\State $vec\_generalized\_Q = Emb\_sql(relaxation(Q))$
\State $\hat{Q} = rep\_selection(vec\_generalized\_Q)$

\State $\hat{D} = \hat{Q}(D)$
\State $initial\_set = Emb\_tab(variational\_subsampling(\hat{D}))$
\For{each $episode$ during training}

\State $batch\_Q$ = $batch(\hat{Q})$
\State action $a_k$ is sampled based on $p(a_k|s_k,\theta)$, $a_k \in A$
\State $s_{k+1}= s_k$ union $a_k$
\State Reward $r$ is then computed based on $state$ and $a_k$ matching to $batch\_Q$ 
\State Use the reward to calculate loss and update the network
    \If{early stopping (loss)}
    \State break
    \EndIf
\EndFor

\State \textbf{return} $M$

\end{algorithmic}
\end{algorithm}


\subsection{Inference and User's Interaction}\label{subsec: inference}
The inference phase, depicted in Figure \ref{fig:rl_reduction_architecture}(b) and detailed in Algorithm~\ref{alg:inference}, occurs after training the RL model. Using the query workload and database, the model outputs the approximation set. Tuple selection for the approximation set is sequential, where groups of tuples are chosen based on the learned policy obtained during model training.

\paragraph{User Interaction} After the initial approximation set is obtained, the user can issues queries.  For each such query, the system estimates whether the model is likely to contain relevant tuples. This estimation is based on the query's closeness to the training workload seen by the model, using the aforementioned query embeddings and the existing model's performance on the training workload.


If the system deems the query answerable, it provides the answer derived from the approximation set. Otherwise, it queries the original database. As will be seen in Section \ref{sec:exp}, our system attains high accuracy in predicting whether a query is answerable. This accuracy significantly contributes to the overall precision of the results presented to the user, minimizing the likelihood of displaying partial results and potentially introducing bias into their analysis.

\paragraph{Identifying Interest Drift} 
Interest drift is identified when user queries deviate from the initial model training query workload. When three or more queries deviate from the training workload with confidence scores surpassing 0.8, our model initiates a fine-tuning process tailored to the specific characteristics of these queries, addressing $C5$. The parameters outlined above that are used to trigger this process is a deliberate effort to strike a balance between optimizing system performance and minimizing running time costs associated with fine-tuning, as detailed in Section \ref{sec:exp}. 

\eat{\paragraph{Generalizing for Future Queries}
Recall that query vectorization and clustering are leverage to choose query representatives to form the initial RL action space. As users begin interacting with our system through queries we assess whether the new queries align with the distribution of the 
query representatives. If disparities arise, 
we generate additional queries that are akin to the user's requests. Subsequently, we fine-tune the model, selecting a different approximation set that is more aligned with the users' evolving interests.  }


\paragraph{Aggregate Queries} 
Recall that in Section \ref{sec:problem_def}, we discussed how aggregate queries in the query workload are transformed into SPJ queries. However, users may also use aggregates when querying our system. As will be shown in Section \ref{sec:exp}, even though the system was not explicitly trained for aggregate queries, it achieves surprisingly good results. Similar to non-aggregate queries, when the estimator perceives a query to be significantly different from the query workload (after transforming it into a non-aggregate query in a similar manner), it queries the database and provides the user with an accurate answer.

\begin{algorithm}
\caption{\algoname{ASQP-RL Inference}}\label{alg:inference}
\begin{algorithmic}[1]
\Require Database $D$, trained model $M$, $req\_size$- optional
\Ensure Approximation Set $S$
\While{|S| < $req\_size$}
\State \textbf{Action } $a_k$ is sampled based on $p(a| s_k,\theta), a \in A$
\State // $p(a| s_k,\theta)$ is output by the model's $M$ policy $\pi$

\State append the tuples of $a_k$ to $S$
\EndWhile
\State \textbf{return} $S$

\end{algorithmic}
\end{algorithm}

\subsection{Further Improvements}\label{sec:asqp_light}
\paragraph{$\algoname{ASQP-Light}$}  We introduce an enhanced version of the $\algoname{ASQP-RL}$ algorithm named $\algoname{ASQP-Light}$ in this section. While the core of the algorithm remains unaltered, specific adjustments have been made to improve its efficiency. These modifications involve implementing an early-stopping threshold, increasing the learning rate, and reducing the number of queries considered during the training's pre-processing phase. Despite a slight loss of $10\%$ in quality, these adjustments significantly reduce the running time from an hour to just 30 minutes, as demonstrated in Section \ref{sec:exp}. Notably, $\algoname{ASQP-Light}$ maintains its superior performance compared to competitors.

\paragraph*{Adaptive Configuration} Users now have the flexibility to control various factors in our system, offering a range of settings between the lightest version and our best-performing configuration in terms of quality. Considering the user's time constraints and the portion of the training query workload to execute (given as input), our system dynamically determines the optimal setting for the use case. The RL model is then trained accordingly. This adaptive approach optimizes the trade-off between running time and the quality of the calculated approximation set, tailoring our system to the specific needs of users.
 
\paragraph*{Diversity}
While our metric function does not explicitly include diversity, our chosen RL solution incorporates an exploration policy. This inherent feature leads to a diverse solution, enriching the variety of query results provided to the user. In Section \ref{sec:exp}, we present experiments comparing the diversity of our solution against other methods, showcasing its superiority over competitors.

\paragraph*{Unknown Query Workloads} \label{subsec:unkown_q}
In Section \ref{sec:problem_def}, we define the problem as optimizing a metric function with respect to a given workload. To address scenarios where the query workload is unknown, we draw inspiration from works employing generated query workloads~\cite{MLQueryOptimization,Dbest}. Our system utilizes statistical information collected from the tables, such as the mean and standard deviation of numerical columns, a sampled set of categorical columns (with repetition to account for popularity of certain values), and standard query templates, to generate query workloads. While the query workload generated this way is adequate, as demonstrated in Section \ref{sec:exp}, we conduct additional query generation alongside user input queries during system interactions to ensure alignment with user interests. During the user's queries on the resulting approximation set, there exists an iterative process of generating new queries that align with the user's interests and possibly fine-tuning the system. This incorporates both the user's queries and the newly generated queries, aiming for better alignment with user preferences.
\section{Reinforcement Learning Model Implementation}\label{sec: rl_in_asqp_rl}

Our agent's policy is governed by a reinforcement learning model. We begin by introducing the actor-critic networks and how proximal policy optimization contributes to the creation of a faster and more suitable approximation set (Section \ref{subsec: actior-critic}). Subsequently, we delve into a discussion about the selected environment and explore alternatives along with their limitations. Notably, this research marks the inaugural application of advanced RL concepts, specifically proximal policy optimization, for tabular data purposes. In this domain, the action space is exceptionally large, and the state representation is highly intricate, posing challenges for adapting RL methods.

\begin{algorithm}
\caption{\algoname{Actor Critic Proximal Policy Optimization}}\label{alg:a3c_ppo}
\begin{algorithmic}[1]
\Require Action space $A$
\Ensure actor critic networks with parameters $\theta$ 
\While{not to the max iterations}
\For{$actor=[1,\dots,N$]}
\State Run policy $\pi_{\theta_{old}}$ in environment for $T$ timesteps 
\State compute advantage estimates $\hat{A_1} \dots \hat{A_{T}}$
\EndFor
\State Optimize surrogate $L$ w.r.t. $\theta$, with $K$ epochs and mini-batch size $M \leq T \cdot N$
\State $\theta_{old} \leftarrow \theta$
\EndWhile
\end{algorithmic}
\end{algorithm}

\subsection{Actor-Critic with Proximal Policy Optimization in ASQP-RL}\label{subsec: actior-critic}
In the $\algoname{ASQP-RL}$ system (discussed in \ref{sec:system_framework}), the agent's policy takes the state representation as input, inferring the optimal action for the next step, a pivotal aspect of the RL model. However, the conventional policy network faces challenges in achieving good performance due to potential high variance in the returned rewards. To address this, the actor-critic method is employed. Additionally, to prevent the selection of many similar tuples, an entropy regularization technique is used to fine-tune the objective function for the selection of diverse tuples.
Typical policy-based reinforcement learning methods, akin to classic RL algorithms like $\algoname{REINFORCE}$ \cite{rlSurvey}, primarily focus on optimizing the parameterized policy concerning the expected long-term reward. Specifically, the objective of the policy network is to maximize $J(\theta)$, defined as follows:
$$J(\theta) = E_{\pi_{\theta}}[R(\tau)] = \sum_{\tau} p(\tau|\theta) \sum_{t=1}^{T}r_t$$

Where $\tau = (s1,a1,\dots S_T, a_T)$, represents a trajectory leading to a selected approximation, i.e. $approx\_set=[a_1,...a_T]$. The aim is to increase the probability of selecting the trajectory $\tau$ given a high reward $R(\tau$). The algorithm $\algoname{REINFORCE}$ a uses gradient ascent to update the parameters $\theta$ in the direction of the gradient:

$$\nabla J(\theta) = E_{\pi_{\theta}}(\sum_{t=0}^T \nabla_\theta \log \pi_{\theta} (a_t|s_t)\sum_{t=1}^{T}r_t)$$

\paragraph*{\textbf{Motivation of actor-critic networks}} The $\algoname{REINFORCE}$ algorithm introduces in high variability in cumulative reward values ($\sum_{t=1}^{T}r_t$) , leading to instability and slow convergence during training \cite{rlIntroduction}. To mitigate this, the actor-critic network strategy is employed. The actor network learns the policy distribution for action selection, while the critic network estimates the baseline, acting as the expected long-term reward under a certain state (explained further in \cite{A3C}).

The forward pass of the actor and critic networks is first introduced, followed by an explanation of how the networks are updated based on the reward and baseline.

\paragraph{\textbf{The actor-critic network}} The actor-critic network comprises two main components: the actor and the critic. The actor employs a policy function $\pi(a_t|s_t; \theta)$, parameterized by $\theta$, to select actions, while the critic estimates action values using a value function $V^{\pi}(s_t;\theta_v)$, parameterized by $\theta_v$. Due to the inability to compute the value function directly, it is estimated using a forward view and a mix of n-step returns to update both the policy and value function. The policy and value function are updated either after every $t_{max}$ actions or when a terminal state is reached. The training stability is enhanced through the use of parallel actor-learners, accumulating updates for improved convergence. Both actor and critic networks are implemented as neural networks (NN), with a large input layer matching the action space's size, followed by smaller fully-connected layers. These layers capture significant patterns between tuples in the database tables. In the critic network, a softmax layer outputs the policy distribution $\pi_\theta(s_t) = \text{Softmax}(\text{NN}(s_t))$, defining the action probabilities. For the critic network, a linear output provides the value function $V(s_t;\theta_V)$. 

Our system trains the RL model with 32 actor and critic networks, asynchronously (as first presented at \cite{nair2015massively}). This parallelization enhances exploration by allowing actors to explore different parts of the environment independently. Different exploration policies are explicitly used in each actor-critic to maximize diversity. This approach stabilizes learning and offers practical benefits such as reduced training time, linearly dependent on the number of parallel actor-learners.


\paragraph{\textbf{Training the actor critic networks}}
Recap that in the $\algoname{REINFORCE}$, the high variance of cumulative rewards previously defined, leads to unstable and inefficient training. To address this, we use the $V$ value as a baseline to be subtracted by the rewards for reducing
the variance which is produced by the critic network and it is
proved that this does not introduce any bias \cite{rlIntroduction}.
To be specific, we use $A(s_t,a_t) = Q^{\pi}(s_t,a_t)-V^{\pi}(s_t)$ (where $Q^{\pi}(s_t,a_t)$ is the expected return starting from state $s_t$, taking action $a_t$, then following policy $\pi$) to replace
the cumulative reward $\sum_{t=1}^{T}r_t$, because it is an estimate of $Q^{\pi}(s_t,a_t)$ and $V^{\pi}(s_t)$ is the baseline. Hence, the gradient becomes,

$$ \nabla J(\theta) \approx \sum_{t=0}^{T}[\nabla_\theta log\pi_{\theta} (a_t|s_t)A(s_t,a_t)]$$

where $A(s_t,a_t)$ can be estimated by $r_t+V^{\pi}(s_{t+1}) - V^{\pi}(s_t)$, named by temporal-difference (TD) error. From another perspective, since the baseline, i.e. the $V$ value, is the expected long-term reward of a state, the \textit{TD} error reflects the  advantages and disadvantages of different actions from the state. Then we discuss how to update the critic network. To estimate the $V$ value accurately, the difference between $r_t+V^{\pi}(s_{t+1})$ and $V^{\pi}(s_t)$ should be as small as possible. Hence, we can find that the \textit{TD} error can also be used to update the critic. To be specific, the loss function should be written as $L_\phi = (r_t+V^{\pi}(s_{t+1}) - V^{\pi}(s_t))^2$.

Concretely, during each episode, the policy and value function estimates are \emph{updated} using the following rule: $\nabla_{\theta'}\log{\pi(a_t|s_t;\theta')}\cdot A(s_t,a_t;\theta,\theta_v)$, where $\theta$ and $\theta_v$ represent the parameters for the policy and value function, respectively. $A(s_t,a_t;\theta,\theta_v)$ is an approximation of the advantage function, which quantifies the advantage of taking action $a_t$ in state $s_t$.

\paragraph{\textbf{Proximal update of the policy}}

We can further improve the quality of the selected tuples by updating more moderately the weights of the network by the adjusting the loss function to contain proximal policy, results in creating \textit{trust region} and clipping the \textit{surrogate objective}, so the updates of the policy will be more moderate, as the maximization of the surrogate objective without a constraint will lead to an excessively large policy update. The major difference is its use of gradient clipping which constrain the policy updates to be within a certain $\epsilon$ range preventing large updates that can cause instability.

$$L^{CLIP}(\theta) = \hat{E_t}[\min(r_t(\theta)\hat{A_t}, clip(r_t(\theta),1-\epsilon, 1+\epsilon)\hat{A_t})$$

were $r_t(\theta)$ denotes the probability ratio $r_t(\theta) = \frac{\pi_{\theta}(a_t |s_t)}{\pi_{\theta_{old}}(a_t |s_t)}$. Maximizing the surrogate objective alone $\hat{E_t}[r_t(\theta)\hat{A_t}]$ without a constraint will lead to excessively large policy update.

\paragraph{\textbf{Further improvements}} To encourage exploration, we incorporates the entropy of the policy $\pi$ into the objective function, mainly adding to each step component $\lambda\nabla_\theta \mathcal{H}(\pi_\theta(\cdot|s_t))$. In this way, the agent is motivated to explore different actions and learn a more diverse and effective policy. 
In addition, to avoid selecting the same tuple multiple times, we implemented action masking \cite{actionMasking} for each algorithm. Action masking enforces a constraint that forbids the agents from selecting certain actions, specifically in our case, those that would result in the repetition of a tuple. 
\eat{Action masking involves restricting certain actions to prevent agents from choosing the same tuple more than once. By applying action masking, the environment enforces a constraint that forbids the agents from selecting certain actions, specifically those that would result in the repetition of a tuple. This prevents redundancy and ensures that each tuple is chosen only once during the decision-making process.}
To try promote diversity among the selected tuples, we employ a regularization function that quantifies the degree of diversity in the chosen subset using a number in $[0,1]$ and add that to the objective function.  This approach is based on the state of the art regularization techniques in ~\cite{regularizationSurvey}.



\paragraph{\textbf{Overall training algorithm}} Algorithm \ref{alg:a3c_ppo} summarizes the above
proposed process.  $\algoname{REINFORCE}$ first initializes the parameters (Line 1). In each iteration (Line 2-7), the system selects a batch of tuples from different tables using the learned policy (Line 3). Then it computes the gradient for each trajectories (Line 4) to update the parameter (Line 6-7). Finally, it outputs a well-trained actor-critic network, with an optimized policy of selecting tuples for approximating non-aggregated queries.

\subsection{Transforming Tabular Domain to RL Environment}
Unlike the domain of computer games, where images and simple discrete information (such as remaining lives or earned points) serve as the main source of state and simple actions such as going one unit in a given direction describe the actions, the tabular domain has seen few algorithms utilizing database tuples as potential actions or tabular data in the environment's state. Our work focuses on constructing an optimal environment specifically designed for our use case and suitable for an RL model. In this section, we elaborate on the chosen environment (briefly introduced previously in Section \ref{subsec: training}) and discuss alternative environments examined in ablation studies during our experiments (Section \ref{sec:exp}).

\paragraph*{Gradual-Set-Learning (GSL) environment}
The selected environment in our system is "gradual-set-learning" (GSL). This environment is initialized with an empty set. An action involves adding several tuples from different tables, drawn from the action space created during the pre-processing phase. After performing an action $a_t$ and reaching a state $S_{t+1}$, the system evaluates the score of the new state using the metric described in Section \ref{sec:problem_def},  $Score(S_{t+1})$,which serves as the reward for the RL agent. An episode concludes either when a terminal state is reached or when the agent reaches $k$ tuples (the $k$'th action.)

\paragraph*{Drop-One (DRP) environment } Another alternative environment for our use case is the "drop-one" (DRP) environment. In this setup, an RL agent concludes an episode when a terminal state is reached or when a predetermined horizon time limit is reached. The environment is initialized as a set of $k$ tuples. Each iteration involves selecting a pair of two actions: first, removing a tuple from the current state, and second, adding another tuple to the state. Alternatively, the agent may choose not to change the current set. After performing an action $a_t$ 
  and reaching a state $S_{t+1}$ (comprising $k$ tuples),  the environment evaluates the score of the new state using the metric described in Section \ref{sec:problem_def}. The reward is calculated as the difference $score(S_{t+1}) - score(S_t)$. The episode concludes when the horizon is reached, with the default set to 100,000-500,000, adjustable by the user.

\paragraph{Discussion} The primary limitation of the DRP method lies in its vulnerability to becoming trapped in a non-optimal approximation set, struggling to eliminate redundant tuples and replace them with more relevant ones. Furthermore, the initialization phase is crucial and unstable, leading to performance variations. Despite exploring various initialization strategies, none completely alleviated these drawbacks. Several alternative environments were also examined, particularly those that combine the GSL and DRP environments. However, none of these environments achieved results as optimal as GSL alone, and additionally, they exhibited larger deviations, indicating less stability. For a more in-depth exploration of potential approaches to tabular data and a comprehensive understanding, further discussions and insights are available in the ablation studies within the experiments section (Section \ref{sec:exp}).

\graphicspath{ {./img/} }

\section{Experiments}\label{sec:exp}

We conducted a comprehensive experimental evaluation of our system, $\algoname{ASQP-RL}$, in this section we present the results of our experiments.

\paragraph{Summary of Findings}
Our experiments consistently showcase the superior performance of subsets chosen by $\algoname{ASQP-RL}$ over alternatives (Figure~\ref{fig:quality_exp}). With a memory size of $15,000$ tuples (approximately $0.1\%$ of the data), our approach achieves an impressive $80\%$ quality, while alternative approaches struggle to reach $60\%$ (Figure 8). Even under extreme memory constraints of only 1000 tuples, $\algoname{ASQP-RL}$ maintains an average quality of $65\%$, surpassing baselines struggling to achieve $40\%$. Our solution demonstrates efficiency, generating a high-quality subset in just one hour compared to $30-40$ hours required by some baselines (Figure~\ref{fig:quality_exp}). This timeframe is practical for offline generation of the approximation set before data exploration sessions. Additionally, in Section \ref{sec:asqp_light}, we introduced a lighter version, $\algoname{ASQP-Light}$, which, according to experimental results, creates the approximation set in 30 minutes with a minor drop in quality. Ablation studies support default parameters and allow for user customization. The impact of different components in our RL system is elucidated, justifying the current system architecture. Preliminary results suggest that $\algoname{ASQP-RL}$ performs well for aggregate queries, obviating the need for additional data storage beyond the proposed subsets.

\subsection{Experimental Setting}
\paragraph*{\underline{\textbf{Datasets}}}
We evaluated the performance of our system over several datasets:
\begin{enumerate}
    \item $\dataset{IMDB-JOB}$ - This dataset~\cite{JOB} contains both data and a query workload. The data is about movies and tv shows, and comes from IMDB. Its size is 34 million tuples.
    \item $\dataset{MAS}$ - The MAS dataset~\cite{MASDb} contains data about researchers and publications. Its size is 600K tuples. The query workload comes from \cite{MASQueries}. 
    \item $\dataset{FLIGHTS}$ - This dataset contains information about flight delays. The data is taken from \cite{flightsIdebench} and queries are generated according to \cite{queryGen}.
\end{enumerate}

\paragraph*{\underline{\textbf{Baselines}}}
To assess the effectiveness of $\algoname{ASQP-RL}$, we conduct comparisons with various baselines designed to address related tasks and potentially offer candidate solutions for the ASQP problem (as detailed in Section \ref{sec:related}). The considered baselines are categorized as naive, database-oriented, and generative models, with each category featuring advanced and diverse representatives:
\paragraph*{Naive baselines:}
\begin{enumerate}

    \item Random sampling ($\algoname{RAN}$) -  Randomly select rows from the large dataset. 
    \item Brute Force ($\algoname{BRT}$) - An algorithm that exhaustively checks different combinations of k tuples to find the optimal solution. It evaluates all possible combinations to determine the best subset of tuples. To manage the computational complexity, a time constraint of 48 hours is imposed on the algorithm. Additionally, parallel processing with 1000 processes is used to expedite the search process by evaluating multiple combinations simultaneously. We then return the best subset found during this process.
    \item Greedy sampling ($\algoname{GRE}$) - In each iteration, take the row that achieves the largest marginal gain with respect to the metric, eliminate this row, and repeat. The running time is limited to 48 hours and 500 processes are used to expedite the process.
    \item Top queried tuples ($\algoname{TOP}$) - Choose a random subset from each query answer. Choose queries that appear in the most queries first, until reaching $k$ tuples.
    \end{enumerate}
\begin{figure}
\resizebox{\textwidth}{!}{

  \centering
  \begin{tabular}{|l|l|l|l|l|l|l|}
    \hline
    \multirow{2}{*}{Baseline} &
      \multicolumn{3}{c|}{IMDB} &
      \multicolumn{3}{c|}{MAS} \\ 
     & Score & setup(m) & QueryAvg(m) &  Score & setup(m) & QueryAvg(m)  \\ \hline
    ASQP-RL & \textbf{0.64$ \pm $0.06}& $60 \pm 7$ &  $0.16 \pm 0.1$ &  \textbf{0.75425 $\pm $0.025} & $30 \pm 2$ & $0.1 \pm 0.12$ \\ \hline
    ASQP-Light & \textbf{0.53$ \pm $0.09}& $32 \pm 2$ &  $0.16 \pm 0.1$ &  \textbf{0.61 $\pm $0.04} & $15 \pm 0.3$ & $0.1 \pm 0.12$ \\ \hline
    VAE &  0.0025$ \pm $0.002 & $1920 \pm 3.5$ &  $5 \pm 0.03$  &   $0.045 \pm 0.001$ & $720 \pm 3$ & $2.5 \pm 0.22$ \\ \hline
    CACH & 0.084$ \pm $0.06 & $330 \pm 12.5$ &  $0.16 \pm 0.1$ &  0.2207 $\pm $0.085 & $34 \pm 1.2$ & $0.1 \pm 0.12$ \\ \hline
    RAN & 0.29$ \pm $0.03& $0.72 \pm 0.08$ &  $0.16 \pm 0.1$ &  0.20275 $\pm $0.034 & $0.68 \pm 0.024$ & $0.1 \pm 0.12$ \\ \hline
    QUIK & 0.343$ \pm $0.04 & $160 \pm 2$ &  $0.16 \pm 0.1$  & 0.25025 $\pm $0.0317 & $60 \pm 3.1$ & $0.1 \pm 0.12$ \\ \hline
    VERD & 0.471$ \pm $0.021& $200 \pm 5$ &  $0.16 \pm 0.1$ &  0.3045 $\pm $0.0467 & $90 \pm 2.3$ & $0.1 \pm 0.12$ \\ \hline
    SKY & 0.347$ \pm $0.001& $1500 \pm 1$ &  $0.16 \pm 0.1$ & 0.33362 $\pm$ 0.001 & $480 \pm 3$ & $0.1 \pm 0.12$  \\ \hline
    BRT & 0.297$ \pm $0.007& $2880 \pm 0.$ &   $0.16 \pm 0.1$ & 0.3975 $\pm $0.001 & $2880 \pm 0.$ & $0.1 \pm 0.12$ \\ \hline
    QRD & 0.3215$ \pm $0.06& $1800 \pm 1$ &  $0.16 \pm 0.1$ & 0.377 $\pm $0.03 & $35 \pm 2$ & $0.1 \pm 0.12$  \\ \hline
    TOP & 0.2707$ \pm $0.0& $338 \pm 21$ &  $0.16 \pm 0.1$ & 0.4592 $\pm $0.0 & $36 \pm 2.3$ & $0.1 \pm 0.12$  \\ \hline
    GRE & N/A& $2880 \pm 0.$ &  N/A & 0.5177 $\pm $0.02 & $2880 \pm 0.$ & $0.1 \pm 0.12$ \\ \hline
    
  \end{tabular}}

  \caption{Quality and Running time}
  \label{fig:quality_exp}
\end{figure}

\paragraph*{Baselines from databases domain:} 
\begin{enumerate}
\setcounter{enumi}{4}
    \item Caching ($\algoname{CACH}$) - Simulates a database's cache \cite{caching} by preserving tuples from the last executed query. This involves managing limited memory to store frequently accessed data, evicting the least recently used (LRU) pages to accommodate new ones.\footnote{In realistic scenarios, the order of query execution may not be neatly separated by user interests, as different users could query the same database with diverse interests, simultaneously. In our experiments, we assume this realistic use case.}
    \item Query result diversification ($\algoname{QRD}$) - An algorithm based on \cite{usingTreeForest}. It follows an iterative approach where it selects the medoids of clusters and then re-assigns the data points to their nearest medoids. 
    \item Skyline ($\algoname{SKY}$) - A technique for summarizing a dataset presented in \cite{skyline}. While a skyline is typically used with numerical values, we extended it to handle categorical columns by comparing two values based on their frequency.
    \item $\algoname{Verdict}$ ($\algoname{VERD}$) - A method for solving AQP proposed in the paper \cite{verdictdb}, that relies on sampling, rewriting queries and adjusting the answers returned from samples.
    \item $\algoname{QuickR}$ ($\algoname{QUIK}$) - Another AQP method, proposed in the paper \cite{quickr}. QuickR relies on sampling and table statistics to keep a catalog of plans and samples and an algorithm for choosing the right samples at the right time.
\end{enumerate}

\textit{Generative model baseline:}
\begin{enumerate}
\setcounter{enumi}{7}
    \item Generative Model ($\algoname{VAE}$) -  We utilized a Variational Autoencoder (VAE) from \cite{VAETABULAR} as a representative of state-of-the-art generative model for approximate query processing for {\em aggregate queries}. 
    The VAE is used to generate fictitious tuples, and queries are executed on the generated data.

\end{enumerate}
\paragraph*{\underline{\textbf{Hyperparmeters}}} 
The actor network we use consists of an input layer followed by 2 fully-connected layers and a softmax layer for getting the logits for choosing the actions. The critic follows a similar architecture except the output is a single number as it is used for estimating the value function. We use a learning rate of $5\cdot10^{-5}$, KL coefficient of $0.2$ and entropy coefficient of $0.001$. All hyperparameters choices are validated in the experiments.
Finally, The problem definition requires us to set values for parameters $F,k$ as is explained in section \ref{sec:problem_def}. By default we set $k=1000$ which is small enough to make the problem interesting but still large enough to make the chosen subset perform well for queries in the
workload. Additionally, we set $F=50$ which is close to real world values (for example the Pandas python library uses a 20 row limit) but still small enough to be informative for a human. We experiment with different values for both these parameters as will be demonstrated later on. Finally, be default the system executes all the queries given during training, this too is changed and tested in the following experiments.
\paragraph*{\underline{\textbf{Environment}}}\label{subsec:environment}
$\algoname{ASQP-RL}$ is implemented in Python 3.9.13 and is an open source library \cite{ourGitLink}.  It can therefore be used, e.g.\  in common Exploratory Data Analysis (EDA) environments such as Jupyter notebooks, to load subsets of any database. We used Ray~\cite{ray} for Reinforcement Learning, OpenAI's Gym~\cite{gym} for the environment interface and Pytorch~\cite{pytorch} for the models. The experiments were run on an Intel Xeon CPU- based server with 24 cores and 96 GB of RAM as well as 2 NVIDIA GeForce RTX 3090 GPUs.

\begin{figure}
\centering    \includegraphics[width=\columnwidth]{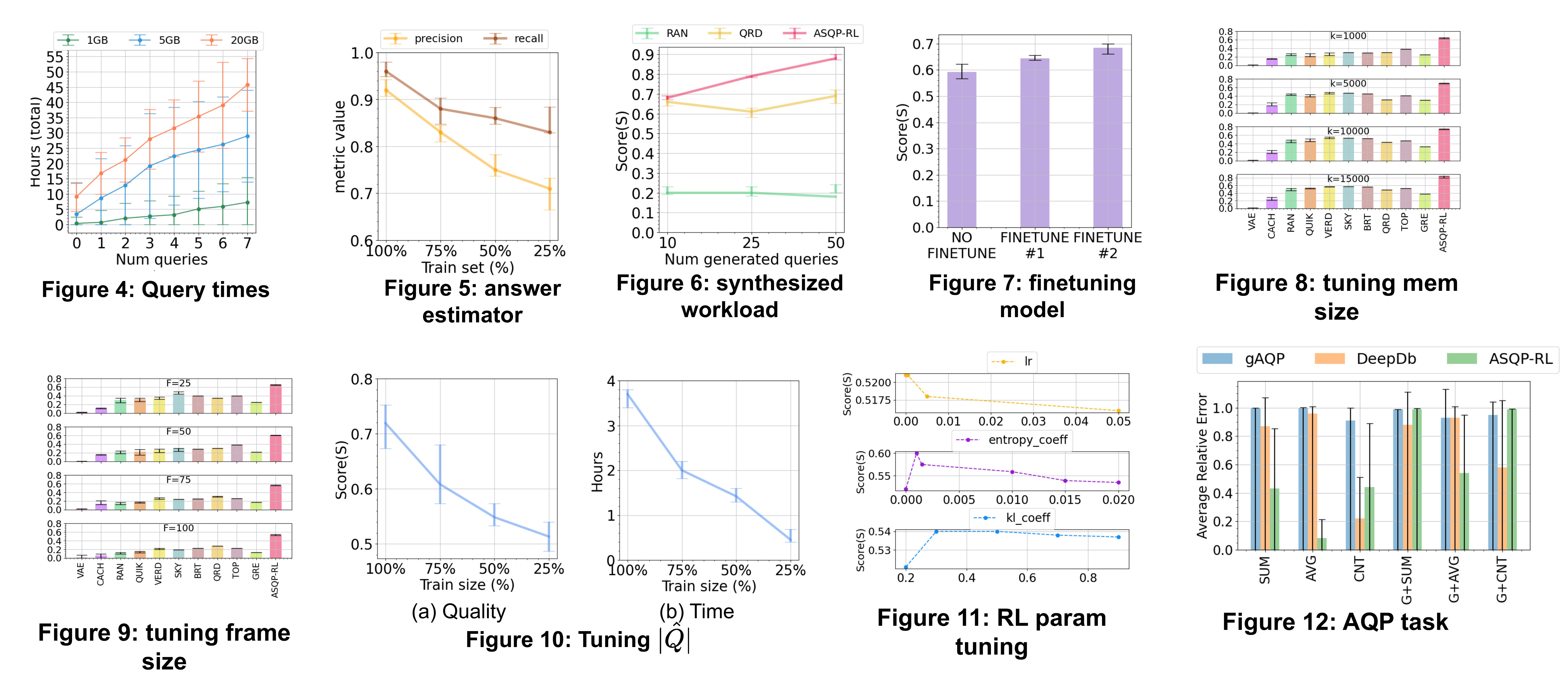}
\end{figure}

\paragraph{\underline{\textbf{Problem Justification}}}
We begin by showing a simple experiment that shows how much were it going to take if we simply query the database directly, in order to heighten the need for solutions such as ours. We take several versions of the $\dataset{IMDB}$ dataset, each one larger then the other (by blowing up the data). We then query the databases with the workload in different orders and average out the time taken for each query - accumulating the results as the number of executed queries grow. Looking at Figure 4 - one can see that even at the relatively small size of 1GB after seven queries on average we have already passed the 5 hour mark.

\subsection{Overall Evaluation} 

In this section, we conduct a comprehensive comparison of our system's performance, focusing on both quality and running time, in contrast to the baselines, across various databases. To facilitate this, we partition the workload $Q$ into a training set ($Q_{train}$) and a test set ($Q_{test}$). For all baselines, the setup utilizes $Q_{train}$ (if necessary), and the evaluation occurs over $Q_{test}$ using the defined approximation quality metric (Equation \ref{eq:metric}, Section \ref{sec:problem_def}). To ensure reliability, we perform multiple train-test partitions, presenting both averaged results and variance.
The presented results align with the default system parameters, and any deviations are investigated through an ablation study, revealing differences in performance.

\paragraph*{\underline{\textbf{Quality Assessment}}}The comparative evaluation of our system's quality against other baselines is succinctly presented in Figure \ref{fig:quality_exp}, emphasizing the $Score$ column, a reflection of our quality metric. This comprehensive examination spans two diverse databases, namely $\dataset{IMDB}$ and $\dataset{MAS}$. A stringent time limit of 48 hours was enforced uniformly across all baselines. Our assessment encompassed both datasets, systematically appraising the efficacy of the approximation set chosen by our algorithm vis-à-vis offerings from alternative baselines. It is imperative to underscore that the greedy ($\algoname{GRE}$) and brute-force ($\algoname{BRT}$) baselines failed to conclude their execution within the stipulated 48-hour timeframe, even when executed in a multi-process environment. Consequently, we present the optimal results achieved within the imposed temporal constraints.

As evident in Figure \ref{fig:quality_exp}, our method unequivocally outperforms conventional standard approaches. Notably, the performance of other algorithms diverged when applied to distinct databases. This variance can be ascribed to the relative size and intricacy of the dataset and workload. "Naive" algorithms such as brute-force ($\algoname{BRT}$) and top-k ($\algoname{TOP}$) exhibited diminished performance when confronted with larger and more intricate workloads. Intriguingly, the $\algoname{VAE}$ algorithm garnered a subpar score due to its inability to generate realistic-looking tuples that align with query filters and its failure to generalize over the workload. This accentuates the shortcomings in data generation concerning selection queries. 

Additionally, we present the scores achieved by $\algoname{ASQP-Light}$, an instantiation of $\algoname{ASQP-RL}$ with a reduced training set, equivalent to $25\%$ of its original size, and a high learning rate of $0.1$ (elaborated upon in subsequent discussions). Noteworthy is the observation that this version attains an average score of $57\%$, while concurrently exhibiting improved running time, as detailed in ensuing discussions.

\paragraph*{\underline{\textbf{Efficiency in Running Time}}} Our system's running time is compared with other methods, focusing on both \emph{set-up} and \emph{query execution} time, as delineated in Figure \ref{fig:quality_exp} under the $setup$ and $QueryAvg$ columns, respectively. The former represents the time (in minutes) each method requires to furnish a queryable subset to the user, while the latter signifies the time it takes to answer a given 10 queries from the workload. These measurements derive from the execution of numerous queries, the same across baselines, with presented results being averages. 
$\algoname{ASQP-RL}$ consistently demonstrates competitive performance, exhibiting a brief setup time compared to most baselines. 

Although naive baselines like Random ($\algoname{RAN}$) may consume less time, their quality, as indicated in the $Score$ column, is notably lower. Divergence exists in the operation of some baselines, notably generative approaches like the $\algoname{VAE}$, necessitating additional time to provide an answer once a query is submitted by the user (approximately five minutes, compared to seconds for other baselines). Noteworthy is the failure of the greedy approach to complete within the allotted 48-hour constraint, rendering it unqueryable.
An additional observation pertains to $\algoname{ASQP-Light}$ (presented earlier), showcasing a remarkable enhancement in the setup time of the system, requiring only 32 minutes for the larger and more intricate $\dataset{IMDB}$ dataset. This signifies a noteworthy stride towards efficiency in our system's operational timelines.
 
\paragraph*{\underline{\textbf{Answers Estimation Quality}}} A pivotal aspect of our system during the inference phase revolves around the estimator, which, when presented with a user query, gauges whether the existing approximation set, crafted by the trained model (as expounded in Section \ref{subsec: inference}), can effectively address the query.
To validate the efficacy of the estimator, we conducted a series of experiments, the results of which are elucidated in Figure 5. The experimental setup involved acquiring a set of test and training queries, generating an approximation set from our system, and subsequently querying the estimator about the answerability of each test query. This information is then juxtaposed with the scores assigned to the queries over the approximation set by our metric, as defined in Equation \ref{eq:metric}. Adopting a threshold of $0.5$ and above as "answerable," we calculated recall and precision to quantitatively compare the estimator's responses with real-world values. Impressively, as depicted in the figure, our approach achieves a remarkable $0.95$ recall and $0.90$ precision. Further iterations of this process involved progressively limiting the system's access to training queries in the training phase. As anticipated, this decrement in familiarity with queries impacted the estimator's performance on the test set queries. Nevertheless, even with reduced training queries ($50\%$ utilization), the precision and recall remained noteworthy, attaining $0.75$ precision and $0.85$ recall.

Building upon our estimator, we implemented the full version of our algorithm. In this variant, the actual database is queried whenever the estimator predicts a query to be unanswerable, allowing users the flexibility to decide on a per-query basis whether they are willing to endure the requisite time for a complete answer. The first version, which queries the database for predictions falling below $60\%$ (according to Equation \ref{eq:metric}), attains an average score of $85\%$ (in contrast to an average of $70\%$ in Figure \ref{fig:quality_exp}). As expected, this results in a rise in "QueryAvg" time to approximately 24 minutes due to the necessity of querying the database. In a second iteration, the database is queried for predictions below $80\%$, the average score is $76\%$, with just  a $\sim5$-minute increase in query response time.

\paragraph*{\underline{\textbf{Handling the No Query Workload Use Case}}} To tackle scenarios where the query workload is not initially provided, as elaborated in Section \ref{sec:system_framework}, we evaluate the system's performance using the $\dataset{FLIGHTS}$ dataset (Figure 6). The system commences by generating multiple queries while simultaneously prompting the user to contribute a subset of queries for refining the generated ones. After incorporating the user-provided queries into our system and constructing an approximation set, we assess the quality of the approximated answers concerning the user's queries. As the user iteratively submits additional queries, the system refines its model through fine-tuning, leading to a performance increase. Throughout this experiment, the user initiates five additional queries at each step, aiding the system in generating more pertinent queries aligned with the user's interests and thereby fine-tuning the model on increasingly relevant queries. 

The obtained results highlight a substantial enhancement in answer quality over successive queries, surpassing alternative methods. We compare our approach against Random ($\algoname{RAN}$) and $\algoname{QRD}$, both capable of running without a workload. While $\algoname{RAN}$ exhibits a lack of consideration for data or workload, $\algoname{QRD}$ utilizes inherent data patterns. However, $\algoname{QRD}$ lags behind our method, failing to surpass $70\%$ accuracy, whereas our approach attains an impressive $90\%$.

\paragraph*{\underline{\textbf{Fine-Tuning Importance}}}
As detailed in Section \ref{subsec: inference}, our system performs a fine-tuning process when it captures an \emph{interest drift} in users queries. This is captured by  utilizing the system's estimator to respond to given queries with the previously created approximation set. Figure 7 summarizes the improvement in quality following the fine-tuning process. We start by dividing our workload into three different subtasks using a clustering algorithm on the embedded version of queries, ensuring that the addition of new queries induces an \emph{interest drift}. We then select a test sample from each cluster for querying the system. Initially, we train the system on the first cluster and query it with the corresponding workload. We gradually introduce parts of the test set matching the training workload the system hasn't seen. As anticipated, the estimator flags these queries as unanswerable, prompting the initiation of fine-tuning by providing the training set of the second cluster. This process is repeated, incorporating the remaining training set. The results demonstrate a rapid enhancement in the quality of the created approximation set when fine-tuning is introduced, making it more aligned with the user's interests.

\paragraph*{\underline{\textbf{Diversity Comparison with Other Baselines}}}
Besides the superior quality and reduced running time of our solutions compared to alternative methods, we conducted experiments to assess the diversity in the approximated query answers provided by our system versus those from other baselines. Although a diversity factor was initially considered for addition to the overall score function and later omitted due to reduced performance, we measured result diversity using a standard metric based on pairwise \emph{Jaccard distance} among query answers. A comprehensive evaluation was conducted using the $\dataset{IMDB}$ dataset, drawing inspiration from various diversity metrics in prior research (e.g.,~\cite{wu2016hear,seleznova2020guided, searchResultDiver2010,DiverseUserSelectionYael}). For an average query in the dataset (when executed with LIMIT 100), the diversity in the answers retrieved directly from the database reached $58\%$. In contrast, our solution achieved an average diversity of $52\%$, showcasing a minimum of $14\%$ higher diversity compared to any other baseline. Notably, our approach competes closely with the $\algoname{RAN}$ baseline, which randomly selects tuples from the database but significantly lags behind in terms of the quality of approximation.

\subsection{System Ablation Studies and Parameters Optimization}

\paragraph*{\underline{\textbf{Effects of Memory Size Constraint (k) on Quality}}}
We investigate how the performance of all baselines is influenced by changes in the memory size ($k$). In this experiment, we test for $k$ in the range $[10^3, 5\cdot10^3, 10\cdot10^3, 15\cdot10^3]$. As expected, all tested approaches should show improvement as the size of the subsets increases. The results in Figure 8 demonstrate that our method consistently outperforms other baselines even with increasing memory size. It achieves an average score of $80\%$ at $k=15\cdot 10^3$, which is double the score of the $\algoname{GRE}$ approach and $20\%$ higher than $\algoname{SKY}$ or $\algoname{QRD}$.

\paragraph*{\underline{\textbf{Effects of Frame Size Constraint (F) on Quality}}}
Similar to the previous experiment, we varied the value of the \emph{frame size} ($F$), incrementing it within the range of $[25, 50, 75, 100]$. Increasing the frame size while keeping the memory size fixed makes the problem more challenging, as more tuples are needed for each query, leading to an overall decrease in quality. Despite this added difficulty, $\algoname{ASQP-RL}$ consistently outperforms all other baselines, as depicted in the results presented in Figure 9. For comparison, observe $\algoname{SKY}$, which starts at $40\%$ and decreases to $20\%$, struggling with the heightened complexity. It is important to note that a reasonable value for $F$ should not exceed a few dozen tuples.

\paragraph*{\underline{\textbf{Effects of Training Set Size ($\hat{Q}$) on Quality and Time}}}
As mentioned in Section \ref{subsec:challenges}, our system takes a set of training queries denoted as $\hat{Q}$ and utilizes an embedding model to determine the most significant queries to execute, denoted as $\hat{Q}{\text{train}}$. This improves overall training time, as running queries can be expensive and time-consuming. However, it comes at the cost of quality since exposing the system to fewer queries may impact its ability to accurately answer similar queries from the test set $Q{\text{test}}$ when training concludes. In Figure 10a, we observe the change in quality as the size of $\hat{Q}_{\text{train}}$ decreases, with our method maintaining reasonable quality compared to other baselines. Additionally, in Figure 10b, we see the training time for the same set of experiments, with training reaching a very low threshold of approximately 30 minutes.

\paragraph*{\underline{\textbf{RL Hyper-parameter Tuning}}}
We test the influence of tuning the main hyper-parameters in the  RL algorithms and present here the results in terms of quality. Shown in Figure 11 is the tuning of the entropy coefficient in range $[0, 0.001, 0.0015, 0.01, 0.015, 0.02]$, the learning rate in range $[5\cdot 10^{-5}, 5\cdot 10^{-4}, 5\cdot 10^{-3}, 5\cdot 10^{-2}]$ and the kl coefficient in range $[0.2, 0.3, 0.5, 0.7, 0.9]$. We conclude that the entropy coefficient has a crucial impact on the success of the algorithm and we've set it to 0.001.

\paragraph*{\underline{\textbf{RL Ablation Studies}}} Figure \ref{fig:rl_ablation} summarizes ablation studies conducted on our RL architecture, aiming to validate the contribution of each component to the overall system performance. Different environments, detailed in Section \ref{sec: rl_in_asqp_rl}, were examined, specifically crafted to transform our problem into a tabular RL environment (Environment column). For each environment, we assessed the impact of removing individual components from our agent, such as the actor-critic architecture and loss clipping (-ac and -ppo in the Agent column). The results indicate that the GSL environment is most suitable for our use case. Moreover, the utilization of our actor-critic architecture, along with clip loss, proves crucial for achieving high system performance.

\subsection{Evaluation for Aggregate queries}\label{subsec: AQP}
While our primary focus is on non-aggregate queries rather than aggregates, we conducted experiments to evaluate the performance of our method in handling aggregate queries. This was done for two reasons:
\textbf{(1)} Understanding Distribution: Aggregate queries allow us to assess whether the distribution of data for specific groupings remains consistent after creating the subset of tuples. This is crucial for understanding if our method preserves the relevant data distribution for the user's query workload. \textbf{(2)} Exploration Phase Performance: Aggregate queries are commonly used in the exploration phase of data analysis. Testing our system's performance on aggregate queries helps determine how well it performs in this phase, in addition to handling non-aggregate queries.

For the AQP task, we compared our solution to two recent leading approaches: $\algoname{gAQP}$ \cite{VAETABULAR} and $\algoname{DeepDB}$ \cite{deepdb}. $\algoname{gAQP}$ uses \emph{Variational Autoencoders} to generate tuples and execute AQP queries on the generated data. $\algoname{DeepDB}$, on the other employs \emph{Sum-Product Networks} to estimate the query answers. We utilized the $\dataset{FLIGHTS}$ dataset and a workload of 1000 aggregate queries generated by \cite{flightsIdebench}. These queries were divided into train-test sets for evaluation. For $\algoname{gAQP}$, we recreated the experiments on the $\dataset{FLIGHTS}$  dataset mentioned in \cite{VAETABULAR}, using a memory size of $1\%$.

To evaluate the results, we employed a common metric for AQP tasks known as relative error (used in both \cite{VAETABULAR}, \cite{deepdb}). This metric compares the predicted answer ($a_{pred}$) to the true answer ($a_{truth}$) of the query. For group by queries, the relative error is computed for each group individually and then averaged. In cases where there are missing groups in the output, the relative error is set to 1 to indicate a complete mismatch between the predicted and true answers. This metric allows us to quantitatively assess the accuracy of our system's predictions in comparison to the ground truth values. The relative error is defined as follows:

\begin{equation}
    \text{relative error} = \frac{|a_{pred}-a_{truth}|}{|a_{truth}|}
\end{equation}

In Figure 12, we present the relative errors for different query categories, including queries with sum (G+SUM, SUM), average (G+AVG, AVG), and count (G+CNT, CNT) operators, both with and without group by clauses (respectively).  Notably, none of the existing approaches outperforms our system across all operators. In half of the operators, our system attains the lowest error rate, surpassing state-of-the-art approaches, while in the remaining cases, we exhibit comparable performance to at least one baseline. These preliminary results for the AQP task are promising, especially considering that our method is not optimized for this task. Fine-tuning our method to better suit this task is left for future work.

\begin{figure}
\resizebox{\textwidth}{!}{

  \centering
  \begin{tabular}{|l|l|l|l|l|l|}
    \hline
    \multirow{2}{*}{Environment} 
    & \multirow{2}{*}{Agent} &
      \multicolumn{2}{c|}{IMDB} &
      \multicolumn{2}{c|}{MAS} \\ 
     & & Score & Total Time (min) & Score & Total Time (min)  \\ \hline
    GSL & $\algoname{ASQP-RL}$ &\textbf{0.64$ \pm $0.06} & $60 \pm 7$ & \textbf{0.75425 $\pm$ 0.025} & $30 \pm 2$  \\ \hline
   
  GSL & $\algoname{ASQP-RL}$ - ppo & 0.5360$\pm$0.003 & $47\pm 5.3$ & $0.623\pm0.071$  & $23\pm 3$ \\ \hline
  
        GSL & $\algoname{ASQP-RL}$ - ppo - ac & 0.496$\pm$0.0015& $72\pm 4.3$ & $0.618\pm0.021$  & $44\pm 2.2$ \\ \hline
        
    DRP & $\algoname{ASQP-RL}$ & 0.3646$ \pm $0.33& $180 \pm 12.3$ & 0.455 $\pm $0.0281 & $97 \pm 3.5$  \\ \hline
   
  DRP & $\algoname{ASQP-RL}$ - ppo & $0.3642\pm 0.0042$ & $165\pm 9.3$ & $0.429\pm0.001$  & $83\pm 2.5$ \\ \hline
  
    DRP & $\algoname{ASQP-RL}$ - ppo - ac & 0.40120$\pm$0.016& $203\pm 7.3$ & $0.423\pm0.011$  & $82\pm 2$ \\ \hline
    
        DRP + GSL & $\algoname{ASQP-RL}$ & 0.569$ \pm $0.02& $73 \pm 5.1$ & 0.619 $\pm $0.001 & $49 \pm 3.6$  \\ \hline
   
  DRP + GSL & $\algoname{ASQP-RL}$ - ppo & 0.51$\pm$0.01 & $71\pm 3.3$ & $0.59\pm0.024$  & $43\pm 2.9$ \\ \hline
  
    DRP + GSL & $\algoname{ASQP-RL}$ - ppo - ac & 0.391$\pm$0.012& $82\pm 5.3$ & $0.51\pm0.0015$  & $44\pm 3.5$ \\ \hline

  \end{tabular}}

  \caption{Reinforcement Learning Ablation Study}
  \label{fig:rl_ablation}
\end{figure}

\section{Conclusions and Future Directions}\label{sec:conc}
This paper defines the problem of approximating answers to non-aggregate queries, $ANAQP$, by creating a subset
of the dataset (an approximation set) over which queries can be executed to provide fast, high quality results. Since it is infeasible to solve the problem exactly, we use RL to provide a good approximate solution.  Offline training, completed within 1 hour, enables on-demand calculation of the approximation set. User queries are then executed over the approximation set with significantly reduced processing times (seconds) compared to executing them over the full dataset (tens of minutes). $\algoname{ASQP-RL}$
overcomes numerous challenges, such as the size of the action-space and the need to generalize beyond the known workload.  Experiments show $\algoname{ASQP-RL}$'s efficiency and quality compared to other baselines. They also show that our approach performs surprisingly well for aggregate queries compared to state-of-the-art techniques. Future work includes fine-tuning the approach, for example using more dataset statistics, ensuring diversity in the query results, and more directly addressing aggregate queries.

\printbibliography

@inproceedings{queryGen,
  title={ML-based cross-platform query optimization},
  author={Kaoudi, Zoi and Quian{\'e}-Ruiz, Jorge-Arnulfo and Contreras-Rojas, Bertty and Pardo-Meza, Rodrigo and Troudi, Anis and Chawla, Sanjay},
  booktitle={2020 IEEE 36th International Conference on Data Engineering (ICDE)},
  pages={1489--1500},
  year={2020},
  organization={IEEE}
}

@inproceedings{MASQueries,
  title={LearnShapley: Learning to Predict Rankings of Facts Contribution Based on Query Logs},
  author={Arad, Dana and Deutch, Daniel and Frost, Nave},
  booktitle={Proceedings of the 31st ACM International Conference on Information \& Knowledge Management},
  pages={4788--4792},
year={2022}
}

@article{JOB,
  title={How good are query optimizers, really?},
  author={Leis, Viktor and Gubichev, Andrey and Mirchev, Atanas and Boncz, Peter and Kemper, Alfons and Neumann, Thomas},
  journal={Proceedings of the VLDB Endowment},
  volume={9},
  number={3},
  pages={204--215},
  year={2015},
  publisher={VLDB Endowment}
}

@inproceedings{ray,
  title={Ray: A distributed framework for emerging AI applications},
  author={Moritz, Philipp and Nishihara, Robert and Wang, Stephanie and Tumanov, Alexey and Liaw, Richard and Liang, Eric and Elibol, Melih and Yang, Zongheng and Paul, William and Jordan, Michael I and others},
  booktitle={13th USENIX Symposium on Operating Systems Design and Implementation (OSDI) 18)},
  pages={561--577},
  year={2018}
}

@article{gym,
  title={Openai gym},
  author={Brockman, Greg and Cheung, Vicki and Pettersson, Ludwig and Schneider, Jonas and Schulman, John and Tang, Jie and Zaremba, Wojciech},
  journal={arXiv preprint arXiv:1606.01540},
  year={2016}
}

@book{rlIntroduction,
  title={Reinforcement learning: An introduction},
  author={Sutton, Richard S and Barto, Andrew G},
  year={2018},
  publisher={MIT press}
}

@article{rlSurvey,
  title={Reinforcement learning: A survey},
  author={Kaelbling, Leslie Pack and Littman, Michael L and Moore, Andrew W},
  journal={Journal of artificial intelligence research},
  volume={4},
  pages={237--285},
  year={1996}
}

@article{pytorch,
  title={Pytorch: An imperative style, high-performance deep learning library},
  author={Paszke, Adam and Gross, Sam and Massa, Francisco and Lerer, Adam and Bradbury, James and Chanan, Gregory and Killeen, Trevor and Lin, Zeming and Gimelshein, Natalia and Antiga, Luca and others},
  journal={Advances in neural information processing systems},
  volume={32},
  year={2019}
}

@inproceedings{flightsIdebench,
  title={Idebench: A benchmark for interactive data exploration},
  author={Eichmann, Philipp and Zgraggen, Emanuel and Binnig, Carsten and Kraska, Tim},
  booktitle={Proceedings of the 2020 ACM SIGMOD International Conference on Management of Data},
  pages={1555--1569},
  year={2020}
}

@inproceedings{VAETABULAR,
  title={Approximate query processing for data exploration using deep generative models},
  author={Thirumuruganathan, Saravanan and Hasan, Shohedul and Koudas, Nick and Das, Gautam},
  booktitle={2020 IEEE 36th international conference on data engineering (ICDE)},
  pages={1309--1320},
  year={2020},
  organization={IEEE}
}

@article{deepdb,
  title={Deepdb: Learn from data, not from queries!},
  author={Hilprecht, Benjamin and Schmidt, Andreas and Kulessa, Moritz and Molina, Alejandro and Kersting, Kristian and Binnig, Carsten},
  journal={arXiv preprint arXiv:1909.00607},
  year={2019}
}

@article{maxkcoverHardness,
  title={A Note on Max k-Vertex Cover: Faster FPT-AS, Smaller Approximate Kernel and Improved Approximation},
  author={Manurangsi, Pasin},
  journal={arXiv preprint arXiv:1810.03792},
  year={2018}
}

@inproceedings{covidOlap,
  title={Machine learning and OLAP on big COVID-19 data},
  author={Leung, Carson K and Chen, Yubo and Hoi, Calvin SH and Shang, Siyuan and Cuzzocrea, Alfredo},
  booktitle={2020 IEEE International Conference on Big Data (Big Data)},
  pages={5118--5127},
  year={2020},
  organization={IEEE}
}

@misc{MASDb,
  author = {Microsoft},
  title = {Microsoft academic search},
  url = {http://academic.research.microsoft.com},
    year = 2016

}

@misc{ourGitLink,
  author = {anonymous},
  title = {ASQP Github Repository},
  url = {https://anonymous.4open.science/r/SAQP-C90C/},
  year = 2023
}

@misc{IBMOlapSyntax,
  author = {IBM},
  url = {https://www.ibm.com/topics/tm1}
}

@misc{IBMOlapSetup,
  author = {IBM},
  url = {https://www.ibm.com/docs/sr/planning-analytics/2.0.0?topic=tier-tm1-server-installation}
}

@article{nair2015massively,
  title={Massively parallel methods for deep reinforcement learning},
  author={Nair, Arun and Srinivasan, Praveen and Blackwell, Sam and Alcicek, Cagdas and Fearon, Rory and De Maria, Alessandro and Panneershelvam, Vedavyas and Suleyman, Mustafa and Beattie, Charles and Petersen, Stig and others},
  journal={arXiv preprint arXiv:1507.04296},
  year={2015}
}

@inproceedings{Dbest,
  title={Dbest: Revisiting approximate query processing engines with machine learning models},
  author={Ma, Qingzhi and Triantafillou, Peter},
  booktitle={Proceedings of the 2019 International Conference on Management of Data},
  pages={1553--1570},
  year={2019}
}

@inproceedings{olapSurvey,
  title={A comprehensive survey of OLAP: recent trends},
  author={Nanda, Adhish and Gupta, Swati and Vijrania, Meenu},
  booktitle={2019 3rd International Conference on Electronics, Communication and Aerospace Technology (ICECA)},
  pages={425--430},
  year={2019},
  organization={IEEE}
}

@inproceedings{olapSurvey2,
  title={A survey on OLAP},
  author={Dhanasree, K and Shobabindu, C},
  booktitle={2016 IEEE International Conference on Computational Intelligence and Computing Research (ICCIC)},
  pages={1--9},
  year={2016},
  organization={IEEE}
}

@inproceedings{quickr,
  title={Quickr: Lazily approximating complex adhoc queries in bigdata clusters},
  author={Kandula, Srikanth and Shanbhag, Anil and Vitorovic, Aleksandar and Olma, Matthaios and Grandl, Robert and Chaudhuri, Surajit and Ding, Bolin},
  booktitle={Proceedings of the 2016 international conference on management of data},
  pages={631--646},
  year={2016}
}

@inproceedings{verdictdb,
  title={Verdictdb: Universalizing approximate query processing},
  author={Park, Yongjoo and Mozafari, Barzan and Sorenson, Joseph and Wang, Junhao},
  booktitle={Proceedings of the 2018 International Conference on Management of Data},
  pages={1461--1476},
  year={2018}
}

@inproceedings{A3C,
  title={Asynchronous methods for deep reinforcement learning},
  author={Mnih, Volodymyr and Badia, Adria Puigdomenech and Mirza, Mehdi and Graves, Alex and Lillicrap, Timothy and Harley, Tim and Silver, David and Kavukcuoglu, Koray},
  booktitle={International conference on machine learning},
  pages={1928--1937},
  year={2016},
  organization={PMLR}
}

@article{regularizationSurvey,
  title={A comprehensive survey on regularization strategies in machine learning},
  author={Tian, Yingjie and Zhang, Yuqi},
  journal={Information Fusion},
  volume={80},
  pages={146--166},
  year={2022},
  publisher={Elsevier}
}

@inproceedings{agarwal2013blinkdb,
  title={BlinkDB: queries with bounded errors and bounded response times on very large data},
  author={Agarwal, Sameer and Mozafari, Barzan and Panda, Aurojit and Milner, Henry and Madden, Samuel and Stoica, Ion},
  booktitle={Proceedings of the 8th ACM European Conference on Computer Systems},
  pages={29--42},
  year={2013}
}

@article{apqKraska2017,
author = {Galakatos, Alex and Crotty, Andrew and Zgraggen, Emanuel and Binnig, Carsten and Kraska, Tim},
title = {Revisiting Reuse for Approximate Query Processing},
year = {2017},
issue_date = {June 2017},
publisher = {VLDB Endowment},
volume = {10},
number = {10},
issn = {2150-8097},
url = {https://doi.org/10.14778/3115404.3115418},
doi = {10.14778/3115404.3115418},
journal = {Proc. VLDB Endow.},
month = jun,
pages = {1142–1153},
numpages = {12}
}

@INPROCEEDINGS{VisualizationAwareSampling,
  author={Y. {Park} and M. {Cafarella} and B. {Mozafari}},
  booktitle={ICDE}, 
  title={Visualization-aware sampling for very large databases}, 
  year={2016},
  volume={},
  number={},
  
  }

@article{AQPSurvey,
  title={Approximate query processing: What is new and where to go? A survey on approximate query processing},
  author={Li, Kaiyu and Li, Guoliang},
  journal={Data Science and Engineering},
  volume={3},
  pages={379--397},
  year={2018},
  publisher={Springer}
}

@article{GANTABULARSURVEY,
  title={Relational data synthesis using generative adversarial networks: A design space exploration},
  author={Fan, Ju and Liu, Tongyu and Li, Guoliang and Chen, Junyou and Shen, Yuwei and Du, Xiaoyong},
  journal={arXiv preprint arXiv:2008.12763},
  year={2020}
}

@article{cormode2017data,
  title={Data sketching},
  author={Cormode, Graham},
  journal={Communications of the ACM},
  volume={60},
  number={9},
  pages={48--55},
  year={2017},
  publisher={ACM New York, NY, USA}
}

@inproceedings{DiverseUserSelectionYael,
  author    = {Yael Amsterdamer and
               Oded Goldreich},
  title     = {Diverse User Selection for Opinion Procurement},
  booktitle = {Proceedings of the 23rd International Conference on Extending Database
               Technology, {EDBT} 2020, Copenhagen, Denmark, March 30 - April 02,
               2020},
  pages     = {486--497},
  year      = {2020}
}

@article{usingTreeForest,
  title={Using Trees to Depict a Forest},
  author={B. Liu and H. V. Jagadish},
  journal={Proc. VLDB Endow.},
  year={2009},
  volume={2},
  pages={133-144}
}

@article{seleznova2020guided,
  title={Guided exploration of user groups},
  author={Seleznova, Mariia and Omidvar-Tehrani, Behrooz and Amer-Yahia, Sihem and Simon, Eric},
  journal={PVLDB},
  volume={13},
  number={9},
  pages={1469--1482},
  year={2020}
}

@article{searchResultDiver2010,
author = {Drosou, Marina and Pitoura, Evaggelia},
title = {Search Result Diversification},
year = {2010},
issue_date = {March 2010},
publisher = {Association for Computing Machinery},
address = {New York, NY, USA},
volume = {39},
number = {1},
issn = {0163-5808},
url = {https://doi.org/10.1145/1860702.1860709},
doi = {10.1145/1860702.1860709},
journal = {SIGMOD Rec.},
month = sep,
pages = {41–47},
numpages = {7}
}

@inproceedings{queryResultDivers,
  title={On query result diversification},
  author={Vieira, Marcos R and Razente, Humberto L and Barioni, Maria CN and Hadjieleftheriou, Marios and Srivastava, Divesh and Traina, Caetano and Tsotras, Vassilis J},
  booktitle={2011 IEEE 27th International Conference on Data Engineering},
  pages={1163--1174},
  year={2011},
  organization={IEEE}
}

@article{wu2016hear,
  author    = {Ting Wu and
               Lei Chen and
               Pan Hui and
               Chen Jason Zhang and
               Weikai Li},
  title     = {Hear the Whole Story: Towards the Diversity of Opinion in Crowdsourcing
               Markets},
  journal   = {{PVLDB}},
  volume    = {8},
  number    = {5},
  -pages     = {485--496},
  year      = {2015},
}

@article{skyline,
    author = {Papadias, Dimitris and Tao, Yufei and Fu, Greg and Seeger, Bernhard},
    title = {Progressive Skyline Computation in Database Systems},
    year = {2005},
    issue_date = {March 2005},
    publisher = {Association for Computing Machinery},
    address = {New York, NY, USA},
    volume = {30},
    number = {1},
    issn = {0362-5915},
    url = {https://doi.org/10.1145/1061318.1061320},
    doi = {10.1145/1061318.1061320},
    journal = {ACM Trans. Database Syst.},
    pages = {41–82},
    numpages = {42},
}

@article{viewSelectionMat,
  title={A survey of view selection methods},
  author={Mami, Imene and Bellahsene, Zohra},
  journal={Acm Sigmod Record},
  volume={41},
  number={1},
  pages={20--29},
  year={2012},
  publisher={ACM New York, NY, USA}
}

@inproceedings{caching,
  title={Query processing over relational databases with semantic cache: A survey},
  author={Ahmad, Munir and Qadir, Muhammad Abdul and Sanaullah, Muhammad},
  booktitle={2008 IEEE International Multitopic Conference},
  pages={558--564},
  year={2008},
  organization={IEEE}
}

@article{actionMasking,
  title={A closer look at invalid action masking in policy gradient algorithms},
  author={Huang, Shengyi and Onta{\~n}{\'o}n, Santiago},
  journal={arXiv preprint arXiv:2006.14171},
  year={2020}
}

@article{ur2016big,
  title={Big data reduction methods: a survey},
  author={ur Rehman, Muhammad Habib and Liew, Chee Sun and Abbas, Assad and Jayaraman, Prem Prakash and Wah, Teh Ying and Khan, Samee U},
  journal={Data Science and Engineering},
  volume={1},
  pages={265--284},
  year={2016},
  publisher={Springer}
}

@inproceedings{ATENAUserStudy,
  title={Automatically generating data exploration sessions using deep reinforcement learning},
  author={Bar El, Ori and Milo, Tova and Somech, Amit},
  booktitle={Proceedings of the 2020 ACM SIGMOD International Conference on Management of Data},
  pages={1527--1537},
  year={2020}
}

\end{document}